\def\etal{{\it et. al.}}
\documentclass[twocolumn,showpacs,preprintnumbers,amsmath,amssymb]{revtex4}

\usepackage{graphicx}
\usepackage{dcolumn}
\usepackage{color}
\usepackage{bm}
\begin{document}

\title {Structure and stability of copper clusters : A tight-binding\\
molecular dynamics study}
\author{Mukul Kabir}
\altaffiliation[ Mailing author's e-mail address: ]{mukul@bose.res.in}
\author{Abhijit Mookerjee}
\altaffiliation[e-mail address: ]{abhijit@bose.res.in}
\affiliation{S.N. Bose National Centre for Basic Sciences,
 JD Block, Sector III, Salt Lake City, Kolkata 700098,
India}
\author{ A.K. Bhattacharya}
\affiliation{Department of Engineering, University of Warwick,
Coventry CV47AL, U.K.}
\date{\today}

\def\RMP{{\it Rev. Mod. Phys.}\ }
\def\PR{{\it Phys. Rev.}\ }
\def\PRL{{\it Phys. Rev. Lett.}\ }
\def\JCP{{\it J. Chem. Phys.}\ }
\def\CPL{{\it Chem. Phys. Lett.}\ }

\begin{abstract}
In this paper we propose a tight-binding molecular dynamics with parameters fitted
to first-principles calculations on the smaller clusters and with an environment
correction, to be a powerful technique for studying large transition/noble metal
clusters. In particular, 
the structure and stability of $Cu_n$ clusters for $n=3-55$ are studied
by using this technique.  The results for 
small $Cu_n$ clusters ($n=3-9$) show good agreement with {\it ab initio} calculations and available
experimental results. In the size range $10\le n \le 55$ most of the clusters adopt icosahedral
structure which can be derived from the 13-atom icosahedron, the polyicosahedral 19-, 23-, and
26-atom clusters and the 55-atom icosahedron, by adding or removing atoms. 
However, a local geometrical change from icosahedral to decahedral structure is observed for 
$n = 40-44$ and return to the icosahedral growth pattern is found at $n=45$ which continues.
Electronic ``magic numbers" ($n=2$, $8$, $20$, $34$, $40$) in this regime are correctly
reproduced. Due to electron pairing in HOMOs, even-odd alternation is found. 
A sudden loss of even-odd alternation in second difference of cluster binding energy, 
HOMO-LUMO gap energy and ionization potential is observed in the region $n\sim40$ due to 
structural change there. Interplay between electronic and geometrical structure is found.  

\end{abstract}
\maketitle

\section{\label{sec:level1}Introduction}
The study of clusters has become an increasingly interesting topic of research in 
both physics and chemistry in recent years, since they span the gap between the
microscopic and macroscopic materials \cite{heer,brack}. Metallic clusters play a 
central role in catalysis \cite{valden,kink1,joo,hansen} and nanotechnology \cite{
binns, park1,gittins}. 
Cluaters of coinage metals $Cu$, $Ag$ and $Au$ have been used in a 
wide range of demonstration \cite{valden,kink1,joo,hansen,binns, park1,gittins}. 
The determination of structural and electronic properties
and the growth pattern of coinage metal clusters are of much interest both 
experimentally \cite{katakuse,tigges,apai,balerna,montano,taylor,ho,knickelbein,spasov} 
and theoretically \cite{massobrio,calaminici,akeby,kabir,datta,gda}.
The electronic configuration of the coinage metals are
characterized by a closed $d$ shell and a single $s$ valance electron 
[$Cu: Ar(3d)^{10}(4s)^1$, $Ag: Kr(4d)^{10}(5s)^1$, $Au: Xe(5d)^{10}(6s)^1$].
Due to the presence of single $s$ electrons in the atomic outer shells, the 
noble metal clusters are expected to exhibit certain similarities to the alkali 
metal clusters.  Electronic structure of alkali metal clusters are well described 
by the spherical shell model, which has successfully interpreted the ``magic numbers'' 
in $Na_n$ and $K_n$ clusters  \cite{heer,brack}. 
A number of experimental features of noble metal clusters are also qualitatively well 
described in terms of simple $s$ electron shell model. For instance, the mass abundance 
spectrum of $Cu_n^-$, $Ag_n^-$ and $Au_n^-$ clusters, which reflects the stability of 
clusters, can be explained by the one-electron shell model \cite{katakuse}. 
But some experimental studies \cite{tigges,apai,balerna,montano} indicate that the localized 
$d$ electrons of the noble metals play a significant role for the geometrical and 
electronic structure through the hybridization with more extended valence $s$ electron .
Therefore, it is important to include the contribution of $3d$ electrons and the $s-d$ 
hybridization for $Cu_n$ clusters.

Bare copper clusters in the gas phase have been studied experimentally by Taylor {\it et al.} 
\cite{taylor} and Ho {\it et al.} \cite{ho} using photoelectron spectroscopy (PES). 
Knickelbein measured ionization potentials of neutral copper clusters and found evidence of
electronic shell structure \cite{knickelbein}.
Very recently cationic copper clusters have been studied using threshold collision-induced
dissociation (TCID) by Spasov {\it et al.} \cite{spasov}. 
Copper clusters have been also investigated theoretically by various accurate 
quantum mechanical and chemical approaches.
Massobrio {\it et al.} \cite{massobrio} studied the structures and energetics of 
$Cu_n$ ($n=2,3,4,6,8,10$) within the local density approximation of density functional 
theory (DF-LDA) by using the Car Parinello (CP) method. 
Calaminici {\it et al. } \cite{calaminici} used the linear combination of Gaussian-type 
orbitals density functional (LCGTO-DFT) approach to study $Cu_n$, $Cu_n^-$ and $Cu_n^+$ 
clusters with $n\le5$.  Akeby {\it et al.} \cite{akeby} used the configuration 
interaction (CI) method with an effective 
core potential (ECP) for $n\le10$. In an earlier communication \cite{kabir} 
we studied the small $Cu_n$ clusters for $n\le9$ by using full-potential muffin-tin orbitals
 (FP-LMTO) technique.

Ideally, the sophisticated, quantum chemistry based, first-principles methods predict both
the stable geometries and energetics to a very high degree of accuracy. The practical
problem arises from the fact that for actual implementation these techniques are
limited to small clusters only. None of the methods described above can be implemented
for clusters much larger than $\sim$ 10 atoms, because of prohibitive computational
expense. The aim of this communication is to introduce an semi-empirical method, which
nevertheless retains some of the electronic structure features of the problem. The
empirical parameters are determined from first-principles  calculations for
 small clusters, and corrections introduced for
local environmental corrections in the larger clusters.

In recent years empirical tight-binding molecular dynamics (TB-MD) method has been developed as an alternative 
to {\it ab initio} methods. As compared with {\it ab initio} methods, the parameterized 
tight-binding Hamiltonian reduces the computational cost dramatically. The main problem
with the empirical tight-binding methods has always been the lack of transferability of
its empirical parameters.  We shall describe here a technique that allows us to fit the parameters
of the model from a fully {\it ab initio}, self-consistent local spin-density approximation (LSDA) 
based FP-LMTO calculation reported earlier by us \cite{kabir,datta} for the 
smaller clusters  and then make correction
for the new environment for clusters in order to ensure transferability (at least to a degree).

It should be mentioned here that Copper clusters have  also been investigated by other empirical methods. D'Agostino carried
molecular dynamics using a quasi-empirical potential derived from a tight-binding approach for nearly
1300 atoms \cite{gda}. More recently Darby {\it et al.} carried geometry optimization by genetic 
algorithm
using Gupta potential \cite{gupta} for $Cu_n$, $Au_n$ and their alloy clusters in the size range 
$n\le56$ \cite{darby}.
These kinds of empirical atomistic potentials are found to be good to predict ground state geometries 
but can not predict electronic properties such as electronic shell closing effect for $n=$$2$, $8$, 
$20$, $40$, $...$, highest occupied-lowest unoccupied molecular level (HOMO-LUMO) gap energy 
and ionization potential. Our proposed TBMD scheme 
will allow us to extrapolate to the larger clusters to study both the ground state geometries as well as
ground state energetics as a function of  cluster size. 

Menon {\it et al.} have proposed a minimal parameter tight-binding molecular dynamics (TBMD) 
scheme for semiconductors \cite{menon1,menon2,ordejon} and extended the method for  
transition metal ($Ni_n$ and $Fe_n$) clusters \cite{menon3,menon4}. 
Recently Zhao {\it et al.} has applied this method for silver clusters \cite{zhao1}. 
In the present work, we shall introduce a similar TB model for copper.

Using this TBMD method, we shall investigate the
 stable  structures, cohesive energies, relative stabilities,
 HOMO-LUMO gaps  and ionization potentials of $Cu_n$ clusters in the size range $n\le55$. 
We shall indicate the comparison between the present results for small clusters, $n\le9$, with 
those of our previous FP-LMTO calculations and other {\it ab initio} and available 
experimental results. This is essential before we go over to the computationally expensive 
study of larger clusters. 
   
\section{\label{sec:level2}Computational Method}
Menon {\it et al.} introduced a minimal parameter tight-binding molecular dynamics
(TBMD) scheme for transition metal clusters \cite{menon3,menon4}. Here we will describe 
the main ingredients. 
In this tight-binding scheme the total energy $E$ is written
as a sum,
\begin{equation}
E=E_{el}+E_{rep}+E_{bond} .
\end{equation}
$E_{el}$ is the sum of the one-electron energies for the 
occupied states $\epsilon_k$,
\begin{equation}
E_{el}=\sum_k^{occ} \epsilon_k ,
\end{equation}
where the energy eigenvalues $\epsilon_k$ are calculated
by solving the eigenvalue equation 
\begin{equation}
H |\Psi_k\rangle \ = \ \epsilon_k |\Psi_k\rangle,
\end{equation}
where $H$ is the one-electron Hamiltonian and $|\Psi_k\rangle$ is electronic
wave function for $k$th level of the eigenstate. In the TB formulation, the
single particle wavefunctions $|\Psi_k\rangle $ are cast as a linear combination
of orthogonalized basis functions $\Phi_{i\nu}$, in the minimum basis set ($\nu=
s,p_x,p_y,p_z,d_{xy},d_{yz},d_{zx},d_{x^2-y^2},d_{3z^2-r^2}$),
\begin{equation}
|\Psi_k\rangle \ = \ \sum_{i\nu}c^k_{i\nu} |\Phi_{i\nu}\rangle,
\end{equation}
where $i$ labels the ions.

The TB Hamiltonian $H$ is constructed within Slater-Koster scheme \cite{slater}, 
where the diagonal matrix elements are taken to be configuration 
independent and the off-diagonal matrix elements are taken
to have Slater-Koster type angular dependence with respect
to the inter-atomic separation vector $\mathbf r$ and scaled
exponentially with the inter-atomic separation $r$:
\begin{equation}
V_{\lambda,\lambda^{'},\mu} = V_{\lambda,\lambda^{'},\mu}(d) S(l,m,n)exp[-\alpha(r-d)],
\end{equation}
where $d$ is the equilibrium bond length for the fcc bulk copper,  
 $S(l,m,n)$ is the Slater-Koster type function of the 
direction cosines $l, m, n$ of the separation vector ${\bf r}$ and $\alpha$
is an adjustable parameter ($=2/d$) \cite{menon4}.  

The Hamiltonian parameters are determined from the dimensionless universal
parameters $\eta_{\lambda,\lambda^{'},\mu}$ \cite{harrison},
\begin{equation}
V_{\lambda, \lambda^{'}, \mu}(d) = \eta_{\lambda, \lambda^{'}, \mu}\left(\frac
{\hbar^2 r_d^{\tau}}{m d^{\tau+2} }\right),
\end{equation}
Where $r_d$ is characteristic length for the transition metal and the parameter
$\tau = 0$ for $s-s$, $s-p$ and $p-p$ 
interactions, $\tau = 3/2$ for $s-d$ and $p-d$ interactions and $\tau = 3$ 
for $d-d$ interaction.
In Table \ref{tab:table1} we present the parameter $r_d$, the on-site energies $E_s, E_p, 
E_d$ and the universal constants $\eta_{\lambda, \lambda^{'}, \mu}$ for Cu \cite{harrison}.
 According to Ref.\cite{menon3} and Ref.\cite{menon4}, we set $E_s=E_d$ and $E_p$ large enough to 
prevent $p$-orbital mixing \cite{harrison}. 
This choice of our tight-binding parameters reproduces the
band structure of the fcc bulk Cu crystal given by Harrison \cite{harrison}. 

\begin{table}
\caption{\label{tab:table1}Parameter $r_d$, on site energies,$E_s$, $E_p$ and $E_d$
 and the universal constants $\eta_{\lambda, \lambda', \mu}$ for Cu \cite{harrison}.   }
\begin{ruledtabular}
\begin{tabular}{lccc}
Parameter & Value & Parameter & Value \\ 
\hline
$r_d$  & 0.67 \AA & $\eta_{pp\pi}$ & -0.81 \\
$E_s$   & -20.14 eV & $\eta_{sd\sigma}$&  -3.16 \\
 $E_p$  & 100.00 eV & $\eta_{pd\sigma}$& -2.95 \\
 $E_d$  & -20.14 eV &  $\eta_{pd\pi}$ & 1.36 \\
 $\eta_{ss\sigma}$ & -0.48 & $\eta_{dd\sigma}$ & -16.20 \\
$\eta_{sp\sigma}$ & 1.84 & $\eta_{dd\pi}$ & 8.75 \\
$\eta_{pp\sigma}$ & 3.24 & $\eta_{dd\delta}$ & 0.00 \\ 
\end{tabular}
\end{ruledtabular}
\end{table}

The repulsive energy $E_{rep}$ is described by a sum of short-ranged repulsive pair potentials, 
$\phi_{ij}$, which scaled exponentially with inter-atomic distance,
\begin{eqnarray}
E_{rep} &=& \sum_i\sum_{j(>i)} \phi_{ij}(r_{ij})
\nonumber\\
&=&  \sum_i\sum_{j(>i)}\phi_0 exp[-\beta(r_{ij}-d)],
\end{eqnarray} 
where $r_{ij}$ is the separation between the atom $i$ and $j$ and $\beta (=4\alpha)$ is a
parameter. $E_{rep}$ contains ion-ion
repulsive interaction and correction to the double counting of the electron-electron repulsion
present in $E_{el}$. The value of $\phi_0$ is fitted to reproduce the
correct experimental bond length of the Cu dimer $ 2.22$ {\AA} \cite{aslund} 
is given in Table\ref{tab:table2}.

The first two terms of the total energy are not sufficient to exactly 
reproduce cohesive energies of dimers through bulk structures. Toma\`nek 
and Schluter \cite{tomanek} introduced a coordination dependent correction term, 
$E_{bond}$, to the total energy, which does not contribute to the force, it is added
to the total energy after the relaxation has been achieved. However, for the metal clusters,
this correction term is significant in distinguishing various isomers for a given 
cluster \cite{menon4}.
\begin{equation}
E_{bond} = - n\left[ a\left(\frac{n_b}{n}\right)^2 + 
b\left(\frac{n_b}{n}\right) + c \right],
\end{equation}
where $n$ and $n_b$ are the number of atoms and total number of bonds of the 
cluster respectively. Number of bonds $n_b$ are evaluated by summing over all bonds according to cut-off
distance $r_c$ and bond length 
\begin{equation}
n_b = \sum_i\left[exp\left(\frac{r_{ij} - r_c}{\bigtriangleup}\right) + 1\right]^{-1}.
\end{equation}

The parameters $a$, $b$ and $c$ in the equation (6) are then calculated by 
fitting the coordination dependent term, $E_{bond}$, to the {\it ab initio}
 results for three
small clusters of different sizes according to the following equation
\begin{equation}
E_{bond} = E_{{\it ab} \hskip 0.2cm {\it initio}} - E_{el} - E_{rep}.
\end{equation}

Thus we have four parameters $\phi_0$, $a$, $b$, and $c$ in this TB model. These parameters 
are once calculated (given in the Table \ref{tab:table2}) for small clusters to 
reproduce known results (whatever experimental or theoretical) and then 
kept fixed for other arbitrary size cluster. To determine the parameters
$a, b $ and $c$ we use the experimental binding energy of Cu dimer 1.03 eV/atom  
\cite{aslund} and the {\it ab initio} FP-LMTO results for $Cu_4$ and $Cu_6$ in 
Ref. \cite{kabir}. For the $Cu_2$ dimer calculated vibrational frequency ($226$ cm$^{-1}$) has
reasonable agreement with experiment \cite{huber} ($265$ cm$^{-1}$).

In molecular dynamics scheme the trajectories $\{R_j(t)\}$ of the ions are determined by the
potential energy surface $E[\{R_j(t)\}]$ corresponding to the total energy of the electronic
system. The force acting on the $i$-th ion is thus given by,
\begin{eqnarray}
{\mathbf F}_i &=& -{\mathbf \nabla}_{R_i} E[\{R_j\}]
\nonumber\\
&=&-{\mathbf \nabla}_{R_i} \left[ \sum_k \left\langle \Psi_k|H|\Psi_k \right\rangle + E_{rep}\right]
\end{eqnarray}

This equation can be further simplified by making use of the Hellmann-Feynman \cite{feynman} theorem
\begin{equation}
{\mathbf F}_i = - \sum_k \left\langle \Psi_k\left|\rule{0mm}{3mm} 
{\mathbf \nabla}_{R_i} H \right|
\Psi_k \right\rangle -  {\mathbf \nabla}_{R_i}\ E_{rep} .
\end{equation} 

The second term in the above equation is the short-ranged repulsive force.
We should note that Pulay correction term does not play any role in any semi-empirical TBMD
 \cite{colombo}.
The reason is twofold. Within TBMD we directly compute the derivative of the
TB Hamiltonian matrix element and  the basis wavefunctions never
appear explicitely, rather they are implicitely contained in the fitted matrix entries.

The motion of the atoms follow a classical behaviour and is governed by the
Newton's law :
\begin{equation}
m \frac{d^2 {\bf R}_i}{dt^2} = {\bf F}_i ,
\end{equation}
where $m$ is the atomic mass. 
\begin{table}
\caption{\label{tab:table2}The adjustable parameters $\phi_0$, $a, b,$
and $c$.  }
\begin{ruledtabular}
\begin{tabular}{cccc}
 $\phi_0 (eV)$ &  a (eV)&   b (eV)  &  c (eV) \\
\hline
 0.034 &  -0.0671 &  1.2375&   -3.0420\\
\end{tabular}
\end{ruledtabular}
\end{table}

\begin{table*}
\caption{\label{tab:table3} Point group (PG) symmetry, cohesive energy per atom, 
difference in cohesive energy per atom $\triangle E$ and average bond length $\langle r \rangle$
of the ground state structure and different isomers for $Cu_n$ clusters with $n\le9$
 obtained from TB calculation and comparison with {\it ab initio} calculations
 \cite{kabir,akeby,massobrio}.
$\triangle E =0.00$ represents the most stable structure for a particular $n$. 
Cohesive energy corresponding 
to the ground state structure in FP-LMTO \cite{kabir}, DF-LDA \cite{massobrio} 
 (in parentheses) calculations and the values from TCID experiment \cite{spasov} are given. 
For $Cu_7$, $C_{3v}(I)$ is the bicapped trigonal
biprism and $C_{3v}(II)$ is the capped octahedron. }
\begin{ruledtabular}
\begin{tabular}{clclccccc}
 Cluster  & PG& \multicolumn{3}{c}{Binding Energy (eV/atom)}  
&\multicolumn{3}{c}{$\triangle E$ (eV/atom)} & $\langle r \rangle$\\
 &Symmetry   &   Present& Theory\footnotemark[1]&Experiment\footnotemark[2] &  Present & 
Theory\footnotemark[3]  &  Theory\footnotemark[4] & (\AA)\\
\hline
 $Cu_3$ & $C_{2v}$      & 1.43& 1.60(1.63)&1.07$\pm$0.12 & 0.00 &      & 0.00 & 2.25\\
        & $D_{3h}$      & 1.32 &      & & 0.11 & 0.06 &      & 2.24\\
        & $D_{\infty h}$& 1.13 &      & & 0.30 & 0.00 &      & 2.24\\
 $Cu_4$ & $D_{2h}$ & 2.00 &2.00(2.09)&1.48$\pm$0.14 & 0.00 & 0.00 & 0.00& 2.23\\
        & $D_{4h}$ & 1.73 &      & & 0.27 & 0.56 &     & 2.22\\
        & $T_d$    & 1.46 &      & & 0.54 & 0.89 &     & 2.24 \\
$Cu_5$  & $C_{2v}$  & 2.24 &2.19   &1.56$\pm$0.15 & 0.00 & 0.00 &     &2.23  \\
        &$D_{3h} $  & 2.03 &       &            & 0.21 & 0.37 &     &2.38 \\
$Cu_6$ & $C_{5v}$ & 2.54  &2.40(2.49) & 1.73$\pm$0.18& 0.00 & 0.00 & 0.00 & 2.40   \\
       & $C_{2v}$ & 2.40 &            &              &0.14 &      & 0.01 & 2.39\\
       & $O_h $   & 1.98 &            &            &0.56 & 0.87 & 0.04 & 2.41\\
$Cu_7$  & $D_{5h}$ & 2.63 &2.65&1.86$\pm$0.22& 0.00 & 0.00 & &2.41\\
        & $C_{3v}(I)$ & 2.50 &      & &0.13 & 0.32 & &2.63\\
        & $C_{3v}(II)$ & 2.30 &      && 0.33 &      & &2.45\\
$Cu_8$  & $C_s$    & 2.87 &2.73(2.84)&2.00$\pm$0.23& 0.00 & & 0.20 &2.41\\
        & $O_h $   & 2.64 &          &           & 0.23 & &      &2.61\\
        & $D_{2d}$ & 2.57 &          &           & 0.30 & & 0.00 &2.59\\
        & $T_d$    & 2.51 &          &           & 0.36 & & 0.15 &2.39\\
$Cu_9$  & $C_{2}$ & 2.87 &2.80  && 0.00 & & & 2.44\\
        & $C_{2v}$ & 2.84 &      && 0.03 & & & 2.59\\
        & $C_s   $ & 2.60 &      && 0.27 & & & 2.41\\
\end{tabular}
\end{ruledtabular}
\footnotetext[1]{From Kabir {\it et al.} (Ref.\cite{kabir}) and Massobrio {\it et al.} 
(Ref.\cite{massobrio}).}
\footnotetext[2]{Calculated from Spasov {\it et al.} (Ref.\cite{spasov}).}
\footnotetext[3]{From Akeby {\it et al.} (Ref.\cite{akeby}).}
\footnotetext[4]{From Massobrio {\it et al.} (Ref.\cite{massobrio}).}
\end{table*}

For numerical simulation of Newtonian dynamics,  we use the velocity Verlet 
molecular dynamics method \cite{swope} for updating the atomic coordinates, which is given by,
\begin{equation}
{\mathbf R}_i(t+\delta t)= {\mathbf R}_i(t)+ {\mathbf V}_i(t)
\delta t +\frac{1}{2m} {\mathbf F}_i(t)(\delta t)^2 ,
\end{equation}
where the velocity $\mathbf V_i$ of the $i$th atom at $t+\delta t$ is
calculated from $ {\mathbf F}_i$ at $t$ and $t+\delta t$ as
\begin{equation}
 {\mathbf V}_i(t+\delta t)={\mathbf V}_i(t)+\frac{1}{2m}
[{\mathbf F}_i(t)+{\mathbf F}_i(t+\delta t)]\delta t .
\end{equation}

At this stage most authors carry out either dissipative dynamics or free
dynamics with feedback \cite{methfessel}. The reason for this is as
follows : for numerical integration of Newton's equations we have to
choose a {\it finite} time-step $\delta t$. Ideally this should be as
small as possible, but that would require an excessively long time for
locating the global minimum. However, a large choice of $\delta t$
leads to unphysical heating up of the system, leading to instability.
Dissipative dynamics has been suggested as a way of overcoming this.
We add a small extra
friction term carefully ${\mathbf F} \Rightarrow  {\mathbf F} - \gamma m{\mathbf{\dot{R}}}$ \cite{menon4}. In the present calculation $\gamma m = 0.32$ amu/psec, and the time step $\delta t$ is taken to be
$1$ fsec and the total time for molecular dynamics simulation is $\sim$ 100 - 200 psec, depending
upon the cluster size and initial cluster configuration with the several annealing schedule.
Methfessel and Schilfgaarde \cite{methfessel} have also used an alternative technique of  free
dynamics with feedback to overcome the above difficulty.

The results of the molecular dynamics may depend sensitively on the
starting configuration chosen. The final equilibrium configurations often
correspond to local minima of the total energy surface and are metastable
states. For the smaller clusters simulated annealing can lead to the
global minimum. We have found the global minimum configurations of the
smaller clusters by the simulated annealing technique. However, this is often not the case for the larger clusters. Recently more sophisticated techniques like the genetic algorithm has been
proposed \cite{ga1}-\cite{ga4}. We have not tried this out in this communication, but
propose this as an efficient technique for further work.

\section{\label{sec:level2}Results and Discussion}
\subsection{\label{subsec:level2a}Geometry optimization}

\begin{figure*}[t]
{\rotatebox{0}{\resizebox{15cm}{20cm}{\includegraphics{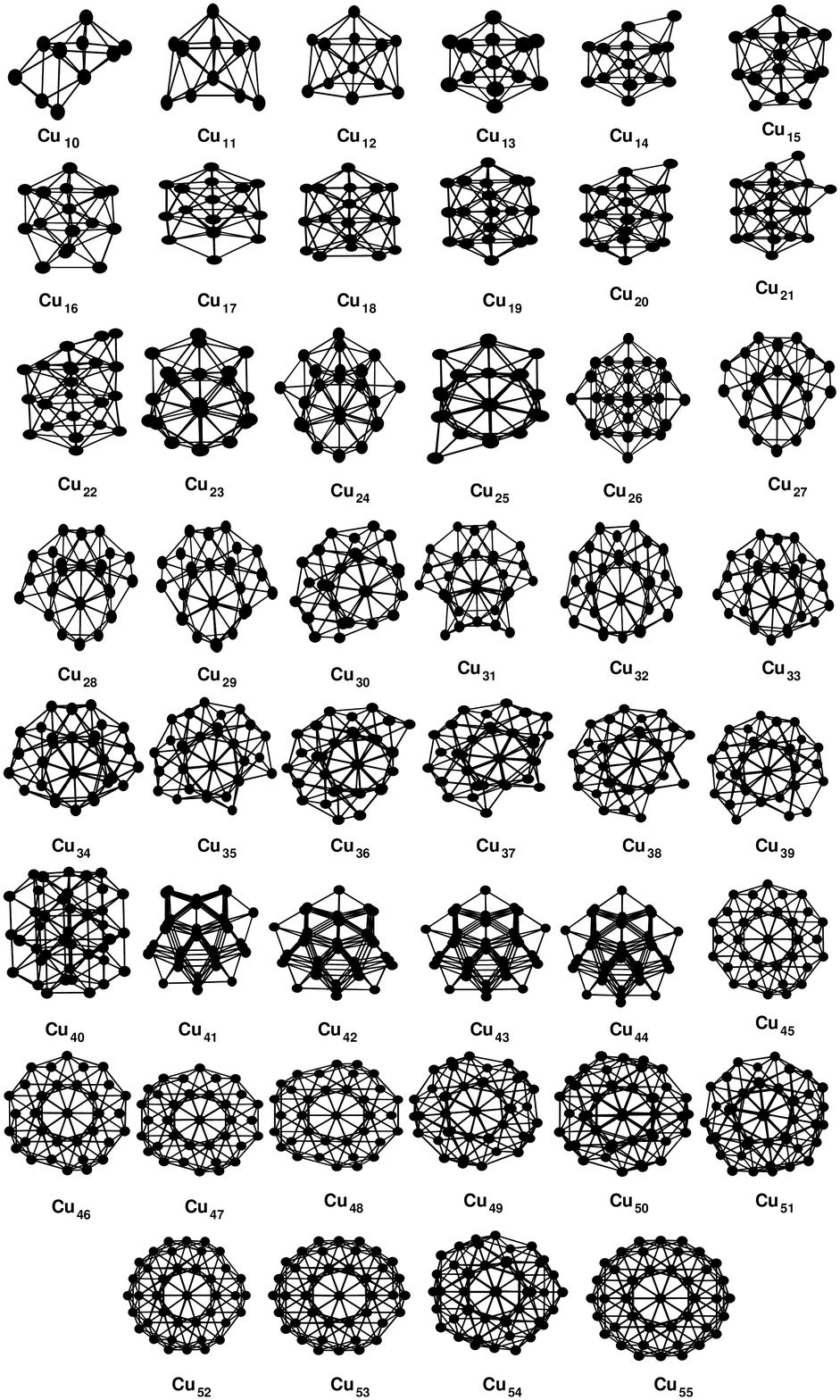}}}}
\caption{\label{fig:structure} Most stable structures for copper clusters with $n=10-55$ atoms. Most of the clusters adopt icosahedral structures except for $n=40-44$, where
the structures are decahedral.}
\end{figure*}

We have applied this TBMD scheme to $Cu_n$ clusters for $n \le 55$. Since the present 
scheme imposes no {\it a priori} symmetry restrictions, we can perform full optimization
of cluster geometries. For small clusters ($n\le9$) we can able to perform a full 
configurational space search to determine the lowest-energy configuration. Here they serve
as a test case for the calculation of larger clusters with $n\ge10$. In Table \ref{tab:table3}
we present a detailed comparison of binding energy per atom, difference in binding energy 
$\triangle E$ and average bond length $\langle r \rangle$ for $n\le9$ with available 
experimental \cite{spasov} and {\it ab initio} \cite{kabir,akeby,massobrio} results. 
We found, in agreement with experimental \cite{ho} and theoretical \cite{massobrio,calaminici,akeby}
 results, very small copper clusters ($Cu_3$, $Cu_4$ and $Cu_5$) prefer planer 
structures. More detailed comparison, with experimental and {\it ab initio} results, can be 
found elsewhere \cite{kabir1}.

From the present results and detailed comparisons with various experimental \cite{ho,spasov}
and {\it ab initio} \cite{massobrio,calaminici,akeby,kabir,bauschlicher,lammers,wang} results 
available, we find
reasonable agreement among this TBMD scheme and {\it ab initio} calculations for small clusters
with $n\le9$ \cite{kabir1}, which allow us to continue the use of this TBMD scheme for larger
clusters with $n\ge10$. For larger clusters ($10\le n \le 55$), due to increasing degrees of 
freedom with cluster size, a full configurational search
is not possible with the available computational resources. Instead, led by the experimental
and theoretical results on small clusters, we examined structures of various symmetries for
each size. Most stable structures for $n=10-55$ atom clusters are given in Fig.\ref{fig:structure}.

In this regime, the structures predicted by this TB model are mainly based on icosahedron. 
The most stable structure of $Cu_7$ is a pentagonal bipyramid ($D_{5h}$ symmetry; see 
Table\ref{tab:table3}), which is the building block for the larger clusters with $n\ge10$.
For $Cu_{10}$, we
found a tricapped pentagonal bipyramid to be the most stable structure. 
 Ground state structures of $Cu_{11}$ and $Cu_{12}$ 
are the uncompleted icosahedron with lack of one and two atoms respectively and a Jahn-Teller
distorted {\it first} closed shell icosahedron is formed at $Cu_{13}$. For $Cu_{13}$, the fcc like
cuboctahedron is less stable than the icosahedron by an energy $0.05$ eV per atom. In agreement 
Lammers and Borstel, on the basis of tight-binding linear muffin-tin orbital calculations, was also
found the icosahedron to be the ground state of $Cu_{13}$, though the difference in energy between 
the icosahedron and the cuboctahedron was calculated to be only $0.2$ eV/atom \cite{lammers}.
The ground state structures for $Cu_{14}$, $Cu_{15}$, $Cu_{16}$, and $Cu_{17}$ are the 13- atom
icosahedron plus one, two, three and four atoms respectively. A double icosahedron is formed
for $Cu_{19}$ ($D_{5h}$ symmetry). This structure has two internal atoms, 12 six-coordinate atoms at 
either end and five eight-coordinate atoms around the waist of the cluster. Based on the structure for 
$Cu_{19}$, the stable $Cu_{18}$ cluster is a double icosahedron minus one of the six-coordinate
atoms located at either end ($C_{5v}$ symmetry). Icosahedral growth continues for $20 \le n \le 55$ 
atom clusters. Polyicosahedral structure in the form of a `` triple icosahedron" ($D_{3h}$ symmetry;
the structure can be viewed as three interpenetrating double icosahedra) is the most stable
structure for $Cu_{23}$ cluster. The next closed shell polyicosahedra is found for $Cu_{26}$ cluster.
Finally, the {\it second} closed shell icosahedron is formed for $Cu_{55}$ which is more stable than the
closed cuboctahedral structure by an energy difference 6.27 eV.  This can be explained in terms of their surface energy. The surface energy of the icosahedral
structure is lower than that of the cuboctahedral structure, because the atoms on the surface of the 
icosahedron are five-coordinate compared to the four-coordinate atoms on the surface of the 
cuboctahedron. In our calculation, exception to the icosahedral growth is found at around $Cu_{40}$. 
The situation regarding geometrical structure in this size range is more complex.
The structures for $n=40-44$ atom clusters are oblate, decahedron like geometries. Return to
the icosahedral structure is found at $n=45$. In the size range $n=40-44$, the structural sequence is
decahedron-icosahedron-cuboctahedron in decreasing order of stability, whereas in the region $n=45-55$,
the structures retain icosahedron-decahedron-cuboctahedron sequence.

This results are in agreement with the experimental study of Winter and co-workers \cite{winter}, where
they found a bare copper cluster mass spectrum recorded with ArF laser ionization shows a sudden
decrease in the ion signal at $Cu_{42}^+$, and from this observation they argued that a change in 
geometrical structure might occur there, though they have not concluded about the nature of this 
geometrical change. 
They also found a dramatic decrease in water binding energy for 
$Cu_{50}$ and $Cu_{51}$, and concluded that this may represent a return to the icosahedral structure
as the closed shell at $Cu_{55}$ is approached.

D'Agostino \cite{gda} confirms our prediction, who performed molecular dynamics
using a tight-binding many-body potential and found that icosahedral structures are prevalent for
clusters containing less than about 1500 atoms. 
Valkealahti and Manninen \cite{valkealahti}, using effective medium theory, also found 
icosahedral structures
are energetically more favourable than the cuboctahedral structures for sizes up to $n\sim2500$ is
consistent with our result: Fig.\ref{fig:compare} shows cuboctahedral structures are least stable 
among the three structures, icosahedron, decahedron and cuboctahedron.
By contrast, Christensen and Jacobsen 
\cite{christensen1} predicted more fcc-like structures in the size range $n=3-29$, in their 
Monte Carlo simulation using an effective medium potential. But they correctly reproduced the 
``magic numbers'' in that regime \cite{christensen1,christensen}. 

These results can be compared with the genetic algorithm study on copper clusters
by Darby {\it et al.} \cite{darby}, using Gupta potential. In agreement with the
present study, Darby {\it et al.} found that most of the clusters in this regime
adopt structures based on icosahedron.  They 
also found exceptions to the icosahedral growth at around $Cu_{40}$, where the structures 
adopt decahedron like geometries (exact numbers are not available in the Ref.\cite{darby}).
But the present study disagree with the genetic
algorithm study in two points. Firstly, for $25$ atom cluster, they found a more
disordered structure, while the present study predict it to be an icosahedron
based structure which can be derived by removing one surface atom from the 26-atom polyicosahedron. 
Finally, they found an fcc-like truncated octahedral structure for $Cu_{38}$. 
Instead, the present study predict the icosahedron based structure to be the ground state, where
this structure is energetically more favourable than the truncated octahedral structure by an
energy $\triangle E = 0.17$ eV/atom. 
Although the genetic algorithm 
search for global minima is more efficient technique than molecular dynamics, 
use of the empirical atomistic potential is the main reason \cite{reason} for this 
kind of disagreement between Darby {\it et al.} and the present study.   

\subsection{\label{subsec:level2b}Binding energies and relative stabilities}

\begin{figure}[b]
{\rotatebox{0}{\resizebox{8cm}{10cm}{\includegraphics{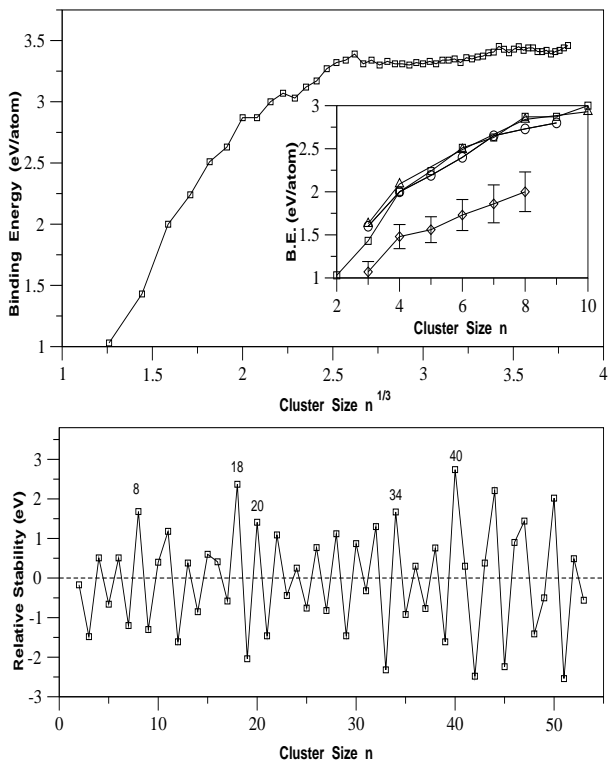}}}}
\caption{\label{fig:bi_st} (Upper panel) Binding energy per atom as a function of cluster size $n^{1/3}$.
Inset of the upper panel represents a comparison of binding energy per atom as a function of cluster
size $n$, among the present TBMD
($\square$), FP-LMTO ($\bigcirc$), DF-LDA ($\triangle$) calculations and experimental 
($\Large{\diamond}$) 
values. (Lower panel) Variation of relative stability $\triangle_2E$ with cluster size $n$. Shell
closing effect at $n=8, 18, 20, 34, 40$ and even-odd alternation up to $n\sim40$ are found. However,
due to geometrical effect this even-odd alternation is disturbed at $n=11, 13$ and $15$.}  
\end{figure}

The computed size dependence of the binding energy per atom for $Cu_n$ clusters
with $n=2-55$ is depicted in Fig.\ref{fig:bi_st} (upper panel). Among all the isomeric geometries
examined for a certain cluster size $n$, the highest cohesive energy has been 
considered for the Fig.\ref{fig:bi_st}. The overall shape of the curve matches the 
anticipated trend: binding energy grows monotonically with increasing the cluster
size. Inset of the Fig.\ref{fig:bi_st} (upper panel) shows the comparison of our calculated binding
energy with the {\it ab initio} \cite{massobrio,kabir} and experimental 
\cite{spasov} results. Experimentally the binding energies of the neutral clusters
were derived from the dissociation energy data of anionic clusters from the TCID 
experiment \cite{spasov} and  using electron affinities from the PES 
experiment \cite{ho}. The inset figure
shows that our calculated binding energies are in good agreement with those from 
DF-LDA \cite{massobrio} and our previous FP-LMTO \cite{kabir} calculations.
However, our binding energies are systematically overestimated, by an energy
$0.53\pm0.12$ to $0.79\pm0.22$, than the experimental binding energies. The LDA
based {\it ab initio} calculations always over-estimate binding energies. This is
a characteristic of the LDA. In the present study, TB parameters have been fitted 
to the {\it ab initio} LDA calculations for very small calculations \cite{kabir}.
It is not surprising therefore that the binding energies are over-estimated. In 
fact, the present results agree well with other LDA based calculations 
\cite{massobrio,kabir}, all of which overestimate the binding energy.  

\begin{figure}[t]
{\rotatebox{0}{\resizebox{8cm}{5cm}{\includegraphics{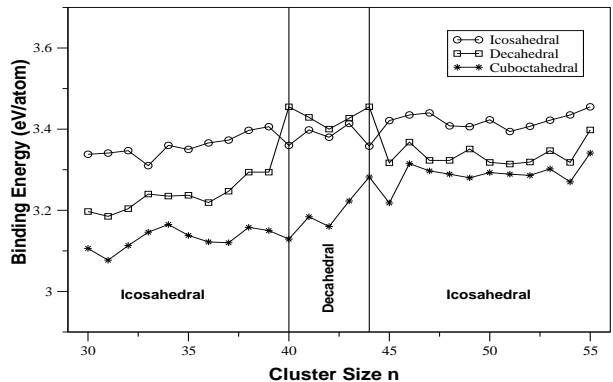}}}}
\caption{\label{fig:compare} Comparison of binding energies per atom as a function of cluster size $n$ among
cuboctahedral, decahedral and icosahedral structures. For the whole region most of the clusters prefer
icosahedral structure. However, a local geometrical change from icosahedral to decahedral structure 
is found for $n=40-44$. 
}  
\end{figure}
In the Fig.\ref{fig:compare}, we compared binding energy per atom for cuboctahedral,
decahedral and icosahedral structures for the clusters containing $n=30-55$ atoms. 
Fig.\ref{fig:compare} shows most the clusters in this size range have icosahedral structures.
However, a local structural change occured for $n=40-44$, where the structures adopt
decahedral structure rather than icosahedral one. Return to the icosahedral growth pattern
is found at $n=45$ and continues up to 55-atom cluster. From the Fig.\ref{fig:compare} it
is clear that among cuboctahedral, decahedral and icosahedral structures, cuboctahedral
structures are least stable than the other two.

The second difference in the binding energy may be calculated as
\begin{equation}
\triangle_2E(n) = E(n+1) + E(n-1) - 2E(n),
\end{equation}
where $E(n)$ represents the total energy for an $n$-atom cluster. 
$\triangle_2E(n)$ represents the relative stability of an $n$-atom cluster with 
respect to its neighbors and can be directly compared to the experimental relative 
abundance : the peaks in $\triangle_2E(n)$ coincide with the discontinuities in the
mass spectra. 
These are plotted in the lower panel of Fig.\ref{fig:bi_st}. We found three major characteristics
in the Fig.\ref{fig:bi_st} (lower panel ). Firstly, even-odd (even $>$ odd) oscillation is found.
This can be explained in terms of electron pairing in HOMOs. Even (odd) clusters have 
an even (odd) number of electrons and the HOMO is doubly (singly) occupied. The electron
 in a doubly occupied HOMO will feel a stronger effective core potential because the 
electron screening is weaker for the electrons in the same orbital than for inner shell
electrons. Thus the binding energy of the valence electron with an even cluster is larger
 than of an odd one. 
This even-odd alternation is prominent up to $n \sim 40$.
Secondly, due to electronic shell or subshell closing, we found particular high peak for
$n=8, 18, 20, 34$ and $40$. Unfortunately, the present study does not show any evidence
of shell closing for $Cu_2$ in $\triangle_2 E(n)$. Finally, the even-odd alternation is reversed
for $n=10-16$ with maxima at $Cu_{11}$, $Cu_{13}$ and $Cu_{15}$, which manifests the geometrical 
effect and therefore there is no peak at $n=14$ due to electronic subshell closing.
Simultaneous appearance of these three features in $\triangle_2 E(n)$ demonstrates the 
interplay between electronic and geometrical structure, which is in agreement with the 
experimental study of Winter {\it et al.} \cite{winter}. They found both jellium-like electronic 
behaviour and icosahedral structure in copper clusters. In an experimental study of mass spectra 
of ionized copper clusters \cite{katakuse}, substantial discontinuities in mass spectra
at $n=3, 9, 21, 35, 41$ for cationic and $n=7, 19, 33, 39$ for anionic clusters as well as
dramatic even-odd alternation are found. From the sudden loss in the even-odd alternation at
$Cu_{42}$ in the KrCl spectrum, Winter {\it et al.} argued about the possible geometrical 
change there. Therefore, we conclude in the section that sudden loss in the $\triangle_2E$ vs $n$
plot (lower panel of Fig.\ref{fig:bi_st}) is due the structural change in that regime. 

Such kind of electronic effects can not be reproduced by empirical atomistic potentials. Darby
{\it et al.} \cite{darby}, using the Gupta potential, found significant peaks at $n=$ 7, 13, 19, 23 and 55 due
to icosahedral (or polyicosahedral) geometries. In the present study, we have found a 
 peak  at $n=13$, but not at the other sizes found by them. However, the stable 
structures predicted by us  are the same : 
the lowest energy structure of $Cu_7$ is a pentagonal bipyramid ($D_{5h}$
 symmetry) ; for $n=13$ and $55$, the structure are the {\it first} and {\it second} closed icosahedral
geometries respectively. Polyicosahedral structures are found for $n=19$ (double icosahedron) and
$n=23$ (triple icosahedron) atom clusters. As the reason,  the present study shows significant high peaks at 
$Cu_8$, $Cu_{18}$ and $Cu_{20}$ due to electronic
shell closing effect and average peaks at $Cu_{22}$ and $Cu_{24}$ due to electron pairing effect. 
At these sizes, the electronic effects dominates over the geometrical effects and consequently
the above peaks cannot observed by Darby \etal.

\subsection{\label{subsec:level2c}HOMO-LUMO gap energies}
\begin{figure}[t]
{\rotatebox{0}{\resizebox{8cm}{5.5cm}{\includegraphics{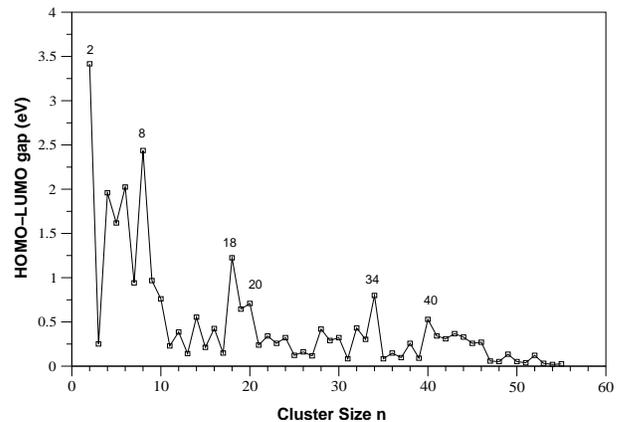}}}}
\caption{\label{fig:homo-lumo} Highest occupied - lowest unoccupied molecular orbital
(HOMO-LUMO) gap energy vs cluster size $n$. Electronic shell closer at $n=2$, $8$, $18$, $20$, $34$, 
$40$ and even-odd alternation are observed. However, sudden loss in even-odd alternation is found
around $n \sim 40$ due to the structural change there.}
\end{figure}

Besides the second difference of the cluster binding energy, a sensitive quantity to probe the 
stability is the highest occupied-lowest unoccupied molecular level (HOMO-LUMO) gap energy. In the 
case of magic clusters shell or subshell closer manifests themselves in particularly large HOMO-LUMO
gap, which was previously demonstrated experimentally \cite{ho,pettiette}. Calculated HOMO-LUMO 
gap energies are plotted in the Fig.\ref{fig:homo-lumo}, where we observed even-odd alternation 
 due to electron pairing effect and particularly
large gap for $n=2$, $8$, $18$, $20$, $34$ and $40$ due to electronic shell and subshell closing. 
However, sudden loss of even-odd alternation 
is found around $n\sim40$ due to the change in the geometrical structure in that region. 
Winter {\it et al.} \cite{winter} also found a sudden loss in even-odd alternation in the KrCl 
spectrum at $Cu_{42}$ and concluded this may coincide with any possible change in the geometrical 
structure there. In fact, Katakuse {\it et al. } \cite{katakuse} observed identical behaviour 
in the mass
spectra of sputtered copped and silver cluster ions : a dramatic loss of even-odd alternation at 
$n=42$, signifying a sudden change to a geometrical structure in which stability, and abundance, 
is less sensitive to electron pairing. Therefore, the sudden loss in the Fig.\ref{fig:homo-lumo}
again confirms the structural change there.
 So, the present study correctly predicts the ``magic numbers''
 in this regime correctly and confirms the experimental prediction : a geometrical change 
(icosahedron  $\rightarrow$ decahedron) is occurring around $n\sim40$.    

\subsection{\label{subsec:level2d}Ionization potentials}
\begin{figure}[t]
{\rotatebox{0}{\resizebox{8cm}{5.5cm}{\includegraphics{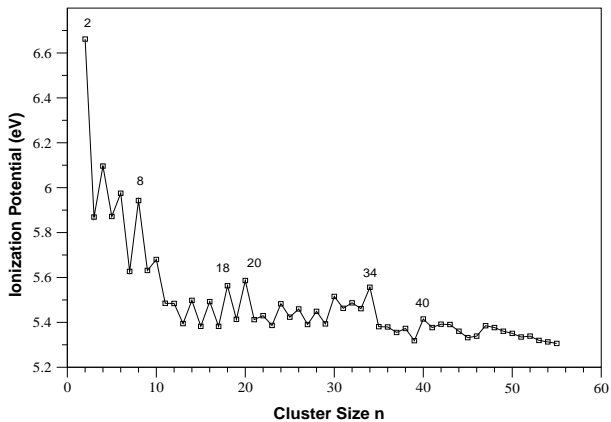}}}}
\caption{\label{fig:ip} Ionization potential vs cluster size $n$. Electronic shell closing
effect and prominant even-odd alternation up to  $n\sim 40 $ are observed. } 
\end{figure}
Within the present TB scheme, we can get a `qualitative' description of the ionization potentials
with cluster size according to Koopmans' theorem. This limitation arises mainly from the choice
of the Slater-Koster (SK) TB parameters and the extent of their transferability \cite{ip}, which
may be improved by the proposed scaling scheme of Cohen, Mehl and Papaconstantopoulos \cite{cohen} .
However, our aim is to get only a qualitative description of ionization potential with cluster size.
Calculated ionization potentials are plotted in the Fig.\ref{fig:ip}. In fact, we observed same pattern
as it was in HOMO-LUMO gap energy {\it vs} cluster size : peaks at $n=2$, $8$, $18$, $20$, $34$, $40$
and even-odd alternation due to the same reasons discussed in the Sec. \ref{subsec:level2b}. 
and Sec. \ref{subsec:level2c}. 
Sudden loss in even-odd alternation around $n\sim40$ is again confirmed from 
the Fig.\ref{fig:ip}, which is due to the geometrical change there.

\section{Conclusion}

Using tight-binding model we calculated ground state geometries, binding energies, second differences
in binding energy, HOMO-LUMO gap energies and ionization potentials for copper clusters in the 
size range $2\le n \le 55$. We have fitted the parameters of the present TB scheme from our 
previous {\it ab inito} calculations \cite{kabir}. For small clusters $n \le 9$, present results 
show good agreement with experimental \cite{ho,spasov} and theoretical 
\cite{massobrio,calaminici,akeby,kabir,bauschlicher,lammers,wang} results, which allow us to go 
over the larger size range, $10 \le n \le 55$.
 
In the size range $10 \le n \le 55$ most of the clusters adopt icosahedral geometry which
can be derived from the 13- atom icosahedron, the polyicosahedral 19-, 23-, and 26-atom 
clusters and 55-atom icosahedron, by adding or removing atoms. However, 
exceptions to the icosahedral growth is found around $n \sim 40$. A local geometrical transition
is found for $n=40-44$- atom clusters.
This is in agreement with the 
prediction of the two experimental studies by Katakuse {\it et al.}\cite{katakuse} and 
Winter {\it et al.}\cite{winter}, where they predicted that a local geometrical transition may 
occur at $n=42$, though their results are not decisive about the nature of this geometrical
change. Present results show that around $n \sim 40 $ structures are changing from icosahedral 
to decahedral structure, where the structural sequence is decahedron-icosahedron-cuboctahedron in 
the decreasing order of stability. Return to the icosahedral growth is found at $n=45$, with the 
sequence icosahedron-decahedron-cuboctahedron in the decreasing order of stability.

As we have fitted the parameters of the present TBMD scheme from LDA based {\it ab inito} 
calculations \cite{kabir}, calculated binding energies are in good agreement with the LDA based
{\it ab initio} calculations but overestimates the same calculated from the TCID experiment  
\cite{spasov}. In the present scheme, the ``magic nimbers'' ($n=2$, $8$, $18$, $20$, $34$ and $40$)
due to electronic shell and subshell closing are correctly reproduced in the studied regime.
Second difference of binding energy, HOMO-LUMO gap energy and ionization potential show even-odd
oscillatory behaviour because of electron pairing in HOMOs in agreement with experiment.
However, a sudden loss in 
even-odd alternation is found around $n \sim 40$ in the variation of second difference in 
binding energy, HOMO-LUMO gap energy and ionization potential with cluster size. This is in 
agreement with the experimental studies \cite{katakuse,winter}. We conclude this is due to 
the geometrical change (icosahedron $\rightarrow$ decahedron) around there. Present results
show that electronic structure can coexist with a fixed atomic packing.

Due to lower computational expense this TBMD scheme, with parameters fitted
to first-principle calculation for the smaller clusters and with an 
environment correction, is a very efficient technique to study larger
clusters, particularly with $n \ge 10$.

\section*{Acknowledgments}
This work is partially supported by the Centre for Catalytic Systems and Materials Engineering,
University of Warwick, U. K.  
The authors are deeply grateful to S. Mukherjee and Luciano Colombo for helpful discussion.


\begin{thebibliography}{99}
\bibitem{heer} W. A. de Heer, \RMP {\bf 65}, 611 (1993).
\bibitem{brack} M. Brack, \RMP {\bf 65}, 677 (1993).
\bibitem{valden} M. Valden, X. Lai and D. W. Goodman, {\it Science} {\bf 281}, 1647 (1998)
\bibitem{kink1} M. B. Kinickelbein, {\it Annu. Rev. Phys. Chem.} {\bf 50} 79 (1999)
\bibitem{joo} S. H. Joo, S. J. Choi, I. Oh, J. Kwak, Z. Liu, O. Terasaki and R. Ryoo, {\it Nature}
 (London) {\bf 412}, 169 (2001).  
\bibitem{hansen} P. L. Hansen, J. B. Wagner, S. Helveg, J. R. Rostrup-Nielsen, B. S. Clausen and H. 
Topsoe, {\it Science} {\bf 295}, 2053 (2002).
\bibitem{binns}C. Binns, {\it Surf. Sci. Rep.} {\bf 44}, 1 (2001).
\bibitem{park1} S. J. Park, T. A. Taton and C. A. Mirkin, {\it Science} {\bf 295}, 1503 (2002).
\bibitem{gittins} D. I. Gittins, D. Bethell, D. J. Schiffrin and R. J. Nicolas, {\it Nature} (London)
 {\bf 408}, 67 (2000).  
\bibitem{katakuse} I. Katakuse, T. Ichihara, Y. Fujita, T. Matsuo, T. Sakurai and H. Matsuda
 {\it Int. J. Mass Spectrom. Ion Proc.} {\bf 67}, 229 (1985); {\bf 74}, 33 (1986).
\bibitem{tigges} J. Tiggesbaumker, L. Koller, K. Meiwes-Broer and A. Liebsch,
\PR  A {\bf 48}, 1749 (1993).
\bibitem{apai} G. Apai, J. F. Hamilton, J. Stohr and A. Thompson, \PRL {\bf 43},
165 (1979).
\bibitem{balerna} A. Balerna, E. Bernicri, P. Piccozi, A. Reale, S. Santucci, E. Burrattini
and S. Mobilio, {\it Surf. Sci.} {\bf 156}, 206 (1985).
\bibitem{montano} P. A. Montano, H. Purdum, G. K. Shenoy, T. I. Morrison and W. Schultze,
{\it Surf. Sci.} {\bf 156}, 216 (1985).
\bibitem{taylor}K. J. Taylor, C. L. Pettiette-Hall, O. Cheshnovsky and R. E. Smalley, 
\JCP {\bf 96}, 3319 (1992).
\bibitem{ho}J. Ho, K. M. Ervin and W. C. Lineberger, \JCP {\bf 93}, 6987 (1990).
\bibitem{knickelbein} M. B. Knickelbein, \CPL {\bf 19}, 2129 (1992).
\bibitem{spasov}V. A. Spasov, T. H. Lee and K. M. Ervin, \JCP {\bf 112}, 1713 (2000).
\bibitem{massobrio} C. Massobrio, A. Pasquarello, R. Car, \CPL,
{\bf 238}, 215 (1995).
\bibitem{calaminici} P. Calaminici, A. M. K\"oster, N. Russo, and D. R. Salahub, \JCP
{\bf 105} 9546 (1996).
\bibitem{akeby} H. Akeby, I. Panas, L. G. M. Pettersson, P. Seigbahn and U. Wahlgren,
\JCP {\bf 94} 5471 (1990). 
\bibitem{kabir} M. Kabir, A. Mookerjee, R. P. Datta, A. Banerjea and A. K. Bhattacharya 
{\it Int. J. Mod. Phys.} B {\bf 17}, 2061 (2003).
\bibitem{datta} R.P. Datta, A. Banerjea, A. Mookerjee and A.K. Bhattacharya, {\it Electronic Structure
of Alloys, Surfaces and Clusters} ed. D.D. Sarma and A. Mookerjee (Taylor and Francis, New York) 348 (2003).
\bibitem{gda} G. D'Agostino, {\it Philos. Mag.} B \ {\bf 68}, 903 (1993).
\bibitem{gupta}R. P. Gupta, \PR B, {\bf 23}, 6265 (1983).
\bibitem{darby}S. Darby, T. V. Mortimer-Jones, R. L. Johnston and C. Roberts, \JCP, {\bf 116}, 1536 
(2002).
\bibitem{menon1}M. Menon and R. E. Allen, \PR B {\bf 33}, 7099 (1986); {\bf 38}, 6196 (1988).
\bibitem{menon2}M. Menon and K. R. Subbaswamy, \PR B {\bf 47}, 12754 (1993); {\bf 50}, 11577 
(1994); {\bf 51}, 17952 (1996).
\bibitem{ordejon}P. Ordej\'on, D. Lebedenko and M. Menon, \PR B, {\bf 50}, 5645 (1994).
\bibitem{menon3} M. Menon, J. Connolly, N. Lathiotaki and A. Andriotis, \PR B, {\bf 50}, 8903 (1994).
\bibitem{menon4} N. Lathiotakis, A. Andriotis, M. Menon and J. Connolly, \JCP,
{\bf 104}, 992 (1996).
\bibitem{zhao1}J. Zhao, Y. Luo and G. Wang, {\it Eur. Phys. J.} D, {\bf 14},
 309 (2001).
\bibitem{slater} J. C. Slater and G. F. Koster, \PR {\bf 94}, 1498 (1954).
\bibitem{harrison} W. A. Harrison, {\it Electronic Structure and the Properties of Solids}
  (Dover,1989).
\bibitem{aslund} N. Aslund, R. F. Barrow, W. G. Richards and D. N. Travis, {\it Ark. Fys.} {\bf 30}
171 (1965).
\bibitem{tomanek} D. Toma\`nek and M. Schluter, \PR B {\bf 36}, 1208 (1987). 
\bibitem{huber}K. P. Huber and G. Herzberg, {\it Molecular Spectra and Molecular Structure}, Vol. IV
(Van Nostrand-Reinhold, New York, 1989).
\bibitem{feynman} R. P. Feynman, \PR {\bf 56}, 340 (1939).
\bibitem{colombo} L. Colombo, ({\it Unpublished}). 
\bibitem{swope}W. C. Swope, H. C. Anderson, P. H. Berenes and K. R. Wilson, \JCP 
{\bf 76}, 637 (1982).
\bibitem{methfessel}M. S. Methfessel and van M. Schilfgaarde, \PR B, {\bf 48}, 4937 (1993); 
{\it Int. J. Mod. Phys.} B, {\bf 7}, 262 (1993); M. S. Methfessel, van M. Schilfgaarde and M. Schaffler,
\PRL {\bf 70}, 29 (1993).
\bibitem{ga1}D. M. Deaven and K. M. Ho, \PRL, {\bf 75}, 288 (1995); D. M. Deaven, N. Tit,
 J. R. Morris and K. M. Ho, \CPL, {\bf 256}, 195 (1996).
\bibitem{ga2}B. Hartke, \CPL, {\bf 240}, 560 (1995).
\bibitem{ga3}Y. H. Luo, J. J. Zhao, S. T. Qiu and G. H. Wang, \PR B, {\bf 59}, 14903 (1999).
\bibitem{ga4}J. A. Niesse and H. R. Mayne, \CPL, {\bf 261}, 576 (1996); \JCP, {\bf 105}, 
 4700 (1996).     
\bibitem{kabir1}M. Kabir, A. Mookerjee and A.K. Bhattacharya, {\it unpublished}.  
\bibitem{bauschlicher} C. W. Bauschlicher, Jr., S. R. Langhoff and H. Partridge, \JCP {\bf 91}, 2412 (1989); C. W. Bauschlicher, Jr., \CPL {\bf 156}, 91 (1989).
\bibitem{lammers} U. Lammers and G. Borstel, \PR B, {\bf 49}, 17360 (1994).
\bibitem{wang} Y. Wang, T. F. George, D. M. Lindsay and A. C. Beri, \JCP, {\bf 86},
3593 (1987); D. M. Lindsay, L. Chu, Y. Wang and T. F. George, {\it ibid.} {\bf 87}, 1685 (1987). 
\bibitem{winter}B. J. Winter, E. K. Parks and S. J. Riley, \JCP, 
{\bf 94}, 8618 (1991).
\bibitem{valkealahti}S. Valkealahti and M. Manninen, \PR B, {\bf 45}, 9459 (1992). 
\bibitem{christensen1} O. B. Christensen and K. W. Jacobsen, {\it J. Phys.: Condens. Matter} \ {\bf 5}, 5591 (1993).
\bibitem{christensen}O. B. Christensen, K. W. Jacobsen, J. K. Norskov and M. Mannien, 
\PRL, {\bf 66}, 2219 (1991).
\bibitem{reason} Even a small variation in the parameters (in particular parameter $q$) in 
the Gupta potential can lead to changes in the global minima. {\it see}- 
K. Michaelian, N. Rendon and I. L. Garz\'on, \PR B \ {\bf 60}, 2000 (1999).
\bibitem{pettiette} C. L. Pettiette, S. H. Yang, M. J. Craycraft, J. Conceicao, R. T. Laaksonen, 
O. Cheshnovsky and R. E. Smalley, \JCP, {\bf 88}, 5377 (1988). 
\bibitem{ip}The set of SK-TB parameters in this scheme implies that within the Koopmans' theorem, 
ionization potentials are approximately equal to the on-site energy $E_s = E_d$, which is usually 
much higher than highest occupied $s$-orbital energy of the free atom. A constant shift is made
in plotting of the Fig.\ref{fig:ip}. {\it See} Ref.\cite{menon4}. 
\bibitem{cohen}R. E. Cohen, H. J. Mehl and D. A. Papaconstantopoulos, \PR B {\bf 50}, 14694, (1994).  


\end{thebibliography}
\end{document}